\documentclass[twocolumn]{aastex62}
\shorttitle{Charged Compact Binary Coalescence Signal}
\shortauthors{Zhang}

\begin{document}


\title{Charged Compact Binary Coalescence Signal and Electromagnetic Counterpart of Plunging BH-NS mergers}


\author{Bing Zhang}
\affil{Department of Physics and Astronomy, University of Nevada Las Vegas, NV 89154, USA}
\email{zhang@physics.unlv.edu}



\begin{abstract}
If at least one of the members of a compact binary coalescence is charged, the inspiral of the two members would generate a Poynting flux with an increasing power, giving rise to a brief electromagnetic counterpart  temporally associated with the chirp signal of the merger (with possibly a small temporal offset), which we term as the {\em charged Compact Binary Coalescence} (cCBC) signal. We develop a general theory of cCBC for any mass and amount of charge for each member. Neutron stars (NSs), as spinning magnets, are guaranteed to be charged, so the cCBC signal should accompany all neutron star mergers. The cCBC signal is clean in a BH-NS merger with a small mass ratio ($q \equiv m_2/m_1 < 0.2$), in which the NS plunges into the BH as a whole, and its luminosity/energy can reach that of a fast radio burst if the NS is Crab-like. The strength of the cCBC signal in Extreme Mass Ratio Inspiral Systems (EMRIs) is also estimated.
\end{abstract}





\section{Introduction}

The discovery of the neutron star - neutron star (NS-NS) merger gravitational wave (GW) event GW170817 \citep{GW170817} and its associated electromagnetic (EM) signals, including the short  GRB 170817A \citep{GW170817/GRB170817A,goldstein17,zhangbb18b}, the ``kilonova'' AT2017gfo \citep{coulter17,villar17}, as well as the broad-band (from radio to X-rays) non-thermal ``afterglow'' \citep{troja17,hallinan17,lyman18,margutti18,piro19}, formally ushered in the era of the GW-led ``multi-messenger'' astrophysics.  

Neutron star mergers (including NS-NS and BH-NS mergers) are widely believed to be bright EM emitters. This is because such systems have plenty of neutron-rich matter outside the horizon of the black hole (if formed) in the merger remnant. Gravitational energy and/or nuclear energy are released as the matter is accreted into the black hole to produce a short GRB \citep{eichler89,narayan92,meszarosrees92}, or as the outgoing ejecta undergo the $r$-process nucleosynthesis and the subsequent $\beta$-decay to power a kilonova \citep{lipaczynski98,metzger10,barnes13,kasen17}. If the post-merger remnant is a long-lived neutron star rather than a black hole, additional energy release is possible with the expense of the spin energy and magnetic energy of the neutron star and its pulsar wind \citep{zhang13,gao13a,yu13,metzger14,piro19}.

For binary black hole (BH-BH) mergers or ``plunging'' BH-NS mergers with a mass ratio $q <0.2$ \citep{shibata09}, it is widely believed that no bright EM emission is expected from the systems, since matter is contained within the horizon of the post-merger black hole. This was why the claimed $\gamma$-ray counterpart, GW150914-GBM, associated with the first BH-BH merger event was regarded controversial \citep{connaughton16,greiner16,connaughton18}, and most theoretical models proposed to interpret the event need to introduce contrived physical conditions to maintain enough matter outside the horizon right after the merger (e.g. \citealt{loeb16,perna16}, c.f. \citealt{kimura17,dai17}).

\cite{zhang16a} suggested that instead of maintaining matter outside the horizon, one may maintain an electromagnetic field in at least one member of the compact binary coalescence (CBC) in order to power an EM counterpart of the CBC. He found that the Poynting-flux luminosity from the system sharply rises right before the merger, which hereafter we refer to the ``charged Compact Binary Coalescence (cCBC) signal''. He hypothesized that at least one BH of the GW150914 system may be significantly charged. He further suggested that with a smaller charge, BH-BH merger systems may account for a fraction of non-repeating fast radio bursts (FRBs), the mysterious millisecond radio bursts at cosmological distances \citep{lorimer07,thornton13}. The charged BH-BH mergers were invoked in several other studies to account for the EM counterparts of BH-BH mergers or FRBs \citep{liebling16,liu16,fraschetti18,levin18,deng18}. Whether BHs can sustain a significant amount of charge is still an open question. \cite{zhang16a} argued that a spinning charged (Kerr-Newmann) black hole can in principle carry a force-free magnetosphere. Like magnetized spinning neutron stars (pulsars), they can sustain a global charge due to the spatial distribution of the charge density demanded to maintain a co-rotating force-free magnetosphere. 

In this paper, we discuss the cCBC signal in general. Since neutron stars are spinning magnets, they are guaranteed to be charged \citep{michel82,petri12}. Even if BHs may not maintain a large enough charge long enough until the merger occurs, NS-NS and BH-NS merger systems should have at least one member (the NS) charged, and the physical process delineated in \cite{zhang16a} should apply.  Given the relatively small charge sustained by neutron stars, the cCBC signal is insignificant in NS-NS merger systems and most BH-NS systems with a relatively large mass ratio $q \equiv m_2/m_1$, where $m_1$ and $m_2 < m_1$ are the masses of the two members in the CBC. This is because the accretion-powered short GRB signal in these systems is orders of magnitude brighter than the cCBC signal. Furthermore, the dynamical ejecta launched during the merger places a significant opacity over a very large solid angle. Certain cCBC signals, e.g. an FRB, would be subject to absorption and may not escape in most solid angles. On the other hand, for BH-NS merger systems with $q < 0.2$ \citep[e.g.][]{shibata09}, the neutron star does not undergo tidal disruption before the merger, which plunges into the black hole entirely. Previously, it has been believed that such systems may not produce EM signals \citep[e.g.][]{bartos13}. Here we suggest that these are ideal systems to observe the cCBC signal. A general theory of cCBC is presented in Sect. \ref{sec:theory}. The strength of the signal for plunging BH-NS systems is discussed in Sect. \ref{sec:BH-NS}. The case of extreme mass ratio inspiral systems is discussed in \ref{sec:EMRI}. The results are summarized in Sect. \ref{sec:summary} with some discussion.

\section{General theory of charged CBC}\label{sec:theory}

Most generally, we consider two members in the CBC that are characterized by $(m_1, \hat q_1)$ and $(m_2, \hat q_2)$, respectively, where $m_i$ is the mass of the member $i = 1,2$, and  $$\hat q_i \equiv Q_i / Q_{c,i}$$ is the absolute charge $Q_i$ divided by its critical charge defined by \citep{zhang16a} $$Q_{c,i} \equiv 2 \sqrt{G} m_i,$$
and $G$ is the gravitational constant.
Besides the total mass $M\equiv m_1+m_2$ and the mass ratio $q = m_2/m_1$, one can define three additional masses from $m_1$ and $m_2$:
\begin{eqnarray}
 {\rm Reduced~mass:} & &M_r = \frac{m_1 m_2}{m_1+m_2}; \\
 {\rm Chirp~mass:} & &M_c = M_r^{3/5} M^{2/5}; \\
 {\rm Horizon~mass:} & &M_h = M_r^{2/5} M^{3/5}.
\end{eqnarray}
The first two masses are commonly used in the gravitational wave community, and the meaning of the third mass ($M_h$) will be introduced shortly.

\subsection{Gravitational wave luminosity}

For easy comparison with the electromagnetic luminosities presented in later subsections, it is informative to write down the gravitational wave (GW) luminosity in several different forms. The first two forms are the standard expressions \citep[e.g.][]{maggiore}
\begin{eqnarray}
 L_{\rm GW} & = & \frac{32}{5} \frac{G^4}{c^5} \frac{M_r^2 M^3}{a^5} f(e), \\
 & = & \frac{32}{5} \frac{c^5}{G} \left(\frac{G M_c \omega_s}{c^3} \right)^{10/3} f(e), \label{eq:LGW2}
\end{eqnarray}
where $c$ is speed of light, $a$ is the semi-major axis, $\omega_s \equiv (GM/a^3)^{1/2}$ is the orbital angular frequency, and
\begin{equation}
	f(e) = \frac{1+\frac{73}{24}e^2+\frac{37}{96}e^4}{(1-e^2)^{7/2}}
\end{equation}
is a correction factor introduced by the orbital eccentricity $e$, which equals unity when $e=0$ (circular orbit). Notice that 
\begin{equation}
 \frac{c^5}{G} \simeq 3.63 \times 10^{59} \ {\rm erg \ s^{-1}}
\end{equation}
is a luminosity unit defined by fundamental constants.

Since $\omega_c$ is directly related to the frequency of gravitational waves, $\omega_{gw}= 2 \omega_s$, the chirp mass $M_c$ can be directly connected to the observed quantities according to Eq.(\ref{eq:LGW2}). For the purpose of our study, the GW luminosity can be expressed in another form
\begin{eqnarray}
 L_{\rm GW} & = & \frac{1}{5} \frac{c^5}{G} \left(\frac{r_s(M_h)}{a}\right)^5 f(e) \nonumber \\
 & = & \frac{1}{5} \frac{c^5}{G} \left(\frac{r_s(M_r)}{a}\right)^2\left(\frac{r_s(M)}{a}\right)^3 f(e), \label{eq:LGW3}
\end{eqnarray}
where 
\begin{equation}
 r_s(M_h) \equiv \frac{2 G M_h}{c^2}
\end{equation}
is the Schwarsczchild (horizon) radius of a non-spinning black hole with mass $M_h$ (hence the name ``horizon'' mass). The advantage of Eq.(\ref{eq:LGW3}) is that one can more straightforwardly track the dependence of $L_{\rm GW}$ on the separation between the two masses, $a$, i.e. $L_{\rm GW} \propto a^{-5}$. For $m_1 = m_2$, one has $M_h = 2^{1/5} m$, so that Eq.(\ref{eq:LGW3}) can be simplified to $L_{\rm GW} = (2/5) ({c^5}/{G}) ({r_s(m)}/{a})^5$.

\subsection{Electric dipole radiation luminosity}

When estimating the luminosity of the cCBC signal, \cite{zhang16a} considered the magnetic dipole radiation power by analogy with pulsars. \cite{deng18} noticed that for a cCBC system, electric dipole radiation is actually more significant. We treat the electric dipole radiation following \cite{deng18} in this subsection.

First, consider that only one member ($m_2$) is charged (with $Q_2$). The electric dipole radiation luminosity is
\begin{eqnarray}
 L_{\rm e,dip,2} & = & \frac{2 Q_2^2 | \ddot {\bf r}_2 |^2} {3 c^3} \nonumber \\
 & = & \frac{8}{3} \hat q_2^2 \frac{G^3 m_1^2 m_2^2} {c^3 a^4} \nonumber \\
 & = & \frac{1}{6} \frac{c^5}{G} \hat q_2^2 \left(\frac{r_s(m_1)}{a} \right)^2 \left(\frac{r_s(m_2)}{a} \right)^2, \label{eq:Ledip1}
\end{eqnarray}
where $r_s(m_1)$ and $r_s(m_2)$ are the Schwarzschild radii for masses $m_1$ and $m_2$, respectively, and $| \ddot {\bf r}_2 | = G m_1/a^2$ is the amplitude of the acceleration of Mass 2. Noticing the symmetric format with respect to the two masses in Eq.(\ref{eq:Ledip1}), it is straightforward to write down the general formula that both masses are charged:
\begin{equation}
 L_{\rm e,dip} = \frac{1}{6} \frac{c^5}{G} (\hat q_1^2+ \hat q_2^2) \left(\frac{r_s(m_1)}{a} \right)^2 \left(\frac{r_s(m_2)}{a} \right)^2.
 \label{eq:Ledip2}
\end{equation}

There are two notes here. First, since the luminosity depends on $\hat q_1^2+\hat q_2^2$, the power enhances no matter whether the two masses have the same or opposite charges. This is generally consistent with the numerical results of \cite{liebling16}. Second, one may write down the ratio
\begin{eqnarray}
 \frac{L_{\rm e,dip}}{L_{\rm GW}}& = &\frac{5}{6} (\hat q_1^2 + \hat q_2^2) \left(\frac{[r_s(m_1)]^2 [r_s(m_2)]^2}{[r_s(M_h)]^4}\right)\left(\frac{a}{r_s(M_h)}\right) \nonumber \\
 & = & \frac{5}{6} (\hat q_1^2 + \hat q_2^2) \left(\frac{M_r}{M}\right)^{2/5} \left(\frac{a}{r_s(M_h)}\right),
\end{eqnarray}
which suggests that the electric dipole luminosity rises more slowly than the GW chirp signal. 

One can also calculate the total dipole radiation energy through integrating luminosity over time. From an initial separation $a$ to the final separation $a_{\rm min}$, one gets
\begin{eqnarray}
 E_{\rm e,dip} & = & \int_a^{a_{\rm min}} da \frac{L_{\rm e,dip}}{\dot a} \nonumber \\
 & = & \frac{5}{24} \ln \left(\frac{a}{a_{\rm min}}\right) (\hat q_1^2+ \hat q_2^2) M_r c^2,
  \label{eq:Eedip}
\end{eqnarray}
where we have used the orbital decay due to the gravitational wave loss  \citep{maggiore}
\begin{equation}
\dot a = -\frac{64}{5} \frac{G^3 M_r M^2}{c^5a^3} f(e), 
\label{eq:adot}
\end{equation}
with $e=0$ (and hence, $f(e)=1$) adopted. Since $e$ also decreases with time for any elliptical orbits, it is reasonable to assume that the orbits are circular when a CBC occurs. Notice that Eq.(\ref{eq:Eedip}) does not converge as $a \rightarrow \infty$. However, since the power increases rapidly at the coalescence, one may mostly care about the energy release during the final orbits. The results are not sensitive to the actual values of $a$ and $a_{\rm min}$ due to the logarithmic factor  $\ln (a/a_{\rm min})$.

\subsection{Magnetic dipole radiation luminosity}

For magnetic dipole radiation, we follow the same procedure of \cite{zhang16a}. The magnetic dipole for the most general case is
\begin{eqnarray}
\mu & = & (\pi/c) I (a/2)^2, \nonumber \\
& = & \sqrt{GMa}(Q_1+Q_2) /8c,
\end{eqnarray}
where $I=(Q_1+Q_2)/P$ is the current, and $P=2\pi(GM)^{-1/2} a^{3/2}$ is the orbital period. The second derivative of $\mu$ reads
\begin{equation}
 \ddot \mu = \frac{\sqrt{GM}}{16 c} (Q_1+Q_2) \left(-\frac{1}{2} a^{-3/2} \dot a^2 +  a^{-1/2} \ddot a\right).
\end{equation}
Again using Eq.(\ref{eq:adot}) with $f(e)=1$, one gets $\ddot a = -(12288/25) (G^6 M_r^2 M^4/a^7 c^{10})$. The magnetic dipole radiation power is then
\begin{eqnarray}
 L_{\rm B,dip} & = & \frac{2 \ddot \mu^2} {3 c^3} \nonumber \\
 & = & \frac{2^{17} \cdot 7^2}{3 \cdot 5^4} \frac{c^5}{G} \left(\frac{\hat q_1 m_1 + \hat q_2 m_2}{M} \right)^2 
 \frac{G^{12} M_r^4 M^{11}}{c^{30} a^{15}} \nonumber \\
 & = & \frac{196}{1875} \frac{c^5}{G} \left(\frac{\hat q_1 m_1 + \hat q_2 m_2}{M} \right)^2 \nonumber \\
& &   \times\left(\frac{r_s(M_r)}{a}\right)^4\left(\frac{r_s(M)}{a}\right)^{11}. \label{eq:LBdip}
\end{eqnarray} 
When $m_1=m_2$, this equation is reduced to Eq.(7) of \cite{zhang16a}.

One can also write down the ratios
\begin{eqnarray}
\frac{L_{\rm B,dip}}{L_{\rm GW}} & = &  \frac{196}{375}  \left(\frac{\hat q_1 m_1 + \hat q_2 m_2}{M} \right)^2 \nonumber \\
&&\left(\frac{r_s(M_r)}{a}\right)^2\left(\frac{r_s(M)}{a}\right)^{8}; \\
\frac{L_{\rm B,dip}}{L_{\rm e,dip}} & = &  \frac{392}{625}  \frac{(\hat q_1 m_1 + \hat q_2 m_2)^2}{(\hat q_1^2+\hat q_2^2) M^2} \nonumber \\
&&\left(\frac{r_s(M_r)}{a}\right)^2 \left(\frac{r_s(M)}{a}\right)^{9},
 \label{eq:LBdip2}
\end{eqnarray}
which suggest that the magnetic dipole radiation power rises much faster than both the GW power and the electric dipole radiation power. At large separations ($a \gg r_s{M_h}$), this term is negligibly small. However, at the merger time, $a \sim r_s(m_1) + r_s(m_2) = r_s(M)$ for BH-BH mergers, $L_{\rm B,dip}$ becomes comparable to $L_{\rm e,dip}$. 

The total magnetic dipole radiation energy can be obtained as
\begin{eqnarray}
 E_{\rm B,dip} & = & \int_a^{a_{\rm min}} da \frac{L_{\rm B,dip}}{\dot a} \nonumber \\
  & = & \frac{49}{4125} (M_r c^2) \left(\frac{\hat q_1 m_1 + \hat q_2 m_2}{M}\right)^2 \nonumber \\
& & \times  \left(\frac{r_s(M_r)}{a_{\rm min}}\right)^2 \left(\frac{r_s(M)}{a_{\rm min}}\right)^{9},
 \label{eq:EBdip1}
\end{eqnarray}
which depends on $a_{\rm min}$. For BH-BH mergers, one has $a_{\rm min} \sim r_s(m_1)+r_s(m_2) = r_s(M)$, so that
\begin{eqnarray}
 E_{\rm B,dip} & = & 
 \frac{49}{4125} (M_r c^2)  \left(\frac{\hat q_1 m_1 + \hat q_2 m_2}{M}\right)^2 \left(\frac{M_r}{M}\right)^2.
 \label{eq:EBdip}
\end{eqnarray}
The ratio between the two total energy components is
\begin{eqnarray}
\frac{E_{\rm B,dip}}{E_{\rm e,dip}} & = & \frac{392}{6875} \frac{(\hat q_1^2 m_1+\hat q_2 m_2)^2}{(\hat q_1^2+\hat q_2^2) M^2 \ln(a/a_{\rm min})} 
\left(\frac{M_r}{M}\right)^2.
\end{eqnarray}
One can see that usually $E_{\rm B,dip} \ll E_{\rm e,dip}$. For $m_1=m_2$, $\hat q_1 = \hat q_2$, this ratio is $\sim 1.8\times 10^{-3} \ln^{-1} (a/a_{\rm min})$.

\subsection{Radiation signature}

The discussion so far does not specify the form of electromagnetic radiation these systems emit. Both electric and magnetic dipole radiations have the frequency of the orbital frequency of the system, which falls around the kHz range for CBCs. This frequency is below the typical plasma frequency $\omega_p = (4\pi n e^2/m)^{1/2} = (5.64 \times 10^4 \ {\rm Hz}) \ n^{1/2}$ for a typical interstellar medium with $n \sim 1 \ {\rm cm^{-3}}$, so the dipole radiations themselves cannot propagate. In reality, the radiation energy is advected in the form of an outgoing Poynting flux dominated outflow. Particle acceleration and subsequent radiation would occur within the outflow with the expense of the Poynting flux energy, so that broad-band radiation (from radio to $\gamma$-rays) is possible. This is certainly the case for spindown-powered pulsars, whose magnetic dipole radiation is released in the form of a pulsar wind and various forms of radiations (e.g. for the Crab pulsar, coherent radio emission, non-thermal $\gamma$-ray and X-ray emission due to photon-pair cascade from the magnetosphere or current sheet outside the light cylinder, and broad band pulsar wind nebula radiation in a much larger scale). Similar physical processes may happen for cCBCs, at least for magnetic dipole radiation, but possibly for electric dipole radiation as well. \cite{zhang16a} discussed the possible mechanisms of converting the magnetic dipole radiation to the magnetospheric coherent radio emission (to power an FRB) and the internal Poynting-flux-dissipation-powered $\gamma$-ray emission (to power a weak GRB).

\section{Electromagnetic counterpart of plunging BH-NS mergers}\label{sec:BH-NS}

For plunging BH-NS mergers, i.e. the mass ratio is required to be $q<0.2$ \citep{shibata09}. The NS is swallowed by the BH as a whole, so that no matter-related EM counterparts (short GRB and kilonova) are expected. 

As has been well known in pulsar theories \citep[e.g.][]{michel82}, rotating, magnetized NSs are charged. For a dipolar magnetic field, even though integrating the Goldreich-Julian spatial charge density distribution ($\rho_{\rm GJ} \sim - ({\bf \Omega \cdot B})/2\pi c$, \citealt{goldreich69}) over the volume contained within the magnetosphere gives no net charge (regardless of the inclination angle), the electric field ${\bf E = - (v \times B)}/c$ has a radial component at the NS surface. Gauss's law gives a net charge contained at the center of the NS \citep{michel82,petri12}
\begin{equation}
Q_{\rm NS} = \frac{1}{3} \frac{\Omega B_p R^3}{c} \cos\alpha,
\end{equation}
where $\alpha$ is the inclination angle between the magnetic and rotational axes of the NS. If the NS is uniformly magnetized, the NS charge is $Q_{\rm NS} = - (\Omega B_p)/(2 \pi c) \cdot (4 \pi/3) R^3 \cos\alpha = - (2/3) (\Omega B_p R^3/c) \cos\alpha$ (notice the opposite sign from the dipole case). Since the cCBC emission power scales with $\hat q^2$, only the absolute value of the charge enters the problem. For the uniformly magnetized NS, the NS dimensionless charge has an absolute value
\begin{equation}
| \hat q_{\rm NS} |  \simeq \frac{\Omega B_p R^3}{3 c \sqrt{G} M_{\rm ns}} \cos\alpha \sim 10^{-7} B_{13} P_{-2}^{-1} R_6^3 M_{1.4}^{-1} \cos\alpha.
\end{equation}
For the Crab pulsar ($B_{13} = 0.8$ and $P_{-2} = 3.3$), one has $\hat q_{\rm Crab} = 2.4 \times 10^{-8} \cos\alpha$.

For a plunging BH-NS merger system with $q=0.2$, assuming that the BH charge is much smaller than the NS charge, one can calculate maximum luminosities and total energies of both electric and magnetic dipole radiations using Eqs. (\ref{eq:Ledip2}), (\ref{eq:Eedip}), (\ref{eq:LBdip}), and (\ref{eq:EBdip}). Notice that at the merger (when the NS touches the BH), one has $a_{\rm min} = r_s(m_1) + 2.4 r_s(m_2)$ (the radius of an NS is about 2.4 times larger than its Schwarzschild radius), we finally get ($a/a_{\rm min}=10$ is adopted)
\begin{eqnarray}
 L_{\rm e,dip,max} & = & (5.0\times 10^{42} \ {\rm erg \ s^{-1}}) \ \hat q_{-7}^2, \nonumber \\
 E_{\rm e,dip} & = & (1.0 \times 10^{40} \ {\rm erg}) \ \hat q_{-7}^2, \nonumber \\
 L_{\rm B,dip,max} & = & (1.7\times 10^{38} \ {\rm erg \ s^{-1}}) \ \hat q_{-7}^2, \nonumber \\
 E_{\rm B,dip} & = & (1.2 \times 10^{34} \ {\rm erg}) \ \hat q_{-7}^2.
\end{eqnarray}
One can see that for $\hat q \sim 10^{-7}$, the electric dipole luminosity/energy reaches that of an FRB assuming isotropic radiation (e.g. \citealt{zhang18b} for a detailed calculation of FRB energetics). {Whether such an EM pulse can be emitted in the GHz frequency range to power an FRB is subject to further studies. In any case, the EM signal is brief and has the right luminosity and energy to potentially power a non-repeating FRB.}

\section{Extreme Mass Ratio Inspiral systems}\label{sec:EMRI}
Besides plunging BH-NS events, another type of system to observe cCBC signals is Extreme Mass Ratio Inspiral (EMRI) systems \citep[e.g.][]{Amaro-Seoane07}, which are possible targets for space gravitational wave detectors such as Laser Interferometer Space Antenna (LISA). For such systems, since $q = m_2/m_1 \ll 1$, one has $M_r \sim m_2$, $M \sim m_1$, and $a_{\rm min} \sim r_s(m_1)$. Equation (\ref{eq:Ledip2}), (\ref{eq:Eedip}), (\ref{eq:LBdip}), and (\ref{eq:EBdip}) can be written as
\begin{eqnarray}
 L_{\rm e,dip,EMRI} & = & (6.1\times 10^{36} \ {\rm erg \ s^{-1}}) \hat q_{-7}^2 q_{-4}^2 \left(\frac{a_{\rm min}}{a} \right)^4; \\
 E_{\rm e,dip,EMRI} & = & (5.2 \times 10^{39} \ {\rm erg}) \hat q_{-7}^2 \ln \left(\frac{a}{a_{\rm min}}\right); \\
  L_{\rm B,dip,EMRI} & = & (3.8\times 10^{20} \ {\rm erg \ s^{-1}}) \hat q_{-7}^2 q_{-4}^6 \left(\frac{a_{\rm min}}{a} \right)^{15}; \\
 E_{\rm B,dip,EMRI} & = & (3.0 \times 10^{22} \ {\rm erg}) \hat q_{-7}^2 q_{-4}^4 \left(\frac{a_{\rm min}} {a}\right)^{11}.
\end{eqnarray}
One can see that magnetic dipole radiation is too weak for any observational interest. If electric dipole radiation can be converted to detectable signals (e.g. radio emission), there might be a possibility of detection by continuing to observe targeted sources over a long period of time. For example, for $q=10^{-4}$ and $a = 2 a_{\rm min}$, the electric dipole radiation luminosity is $\sim 3.8 \times 10^{35} \ {\rm erg \ s^{-1}}$. Even if the luminosity is much lower than that of stellar-mass BH-NS systems, these systems are long lived, and the signal becomes stronger after a long-time integration. If LISA identifies an EMRI source in the sky with the NS very close to the event horizon of the massive black hole, long-term radio monitoring is recommended to detect the possible cCBC signal associated with the system.

\section{Summary and Discussion}\label{sec:summary}

Continuing from a previous investigation \citep{zhang16a}, in this paper, we develop a more general theory for CBC systems with at least one member charged and study the cCBC electromagnetic signal in great detail. The current treatment allows different masses and amounts of charge for the two merging members, and the luminosities and energies due to both electric and magnetic dipole radiations are calculated. The most useful expressions are Eqs. (\ref{eq:Ledip2}), (\ref{eq:Eedip}), (\ref{eq:LBdip}), and (\ref{eq:EBdip}). In general, since $\hat q$ of astrophysical objects is typically $\ll 1$, the cCBC signal is not expected to be very bright. 

Even though it is still an open question whether significantly charged Kerr-Newmann BHs can survive a long enough time, it is well known that NSs are globally charged. So, the cCBC signal should present in NS mergers. This signal is likely non-detectable in NS-NS mergers or BH-NS mergers with $q>0.2$, when plenty of matter is present outside of the horizon of the post-merger black hole.The plunging BH-NS mergers with $q<0.2$ are ideal systems to cleanly observe the cCBC signal. Our estimate suggests that in order to have the cCBC signal reaching a detectable level (e.g. that of an FRB), the NS needs to be young with a relatively strong magnetic field, e.g. Crab-like. This may be possible in young star clusters where BH-NS binaries may form in tight orbits so that the merger occurs within a short period of time when the pulsar is still young. Since the charge of the NS $Q \propto \Omega B_p$, the cCBC signal, if detected, can be used to infer the NS parameter ($\Omega B_p$) before falling into the BH.

Besides the cCBC signal, the distortion of the NS magnetosphere itself near the end of CBC may trigger additional magnetic dissipation in the NS magnetosphere, the strength of which depends on the available magnetic energy. For an order of magnitude estimate, the total dissipated energy would be $\sim (B^2 / 8\pi) R^3 \sim (4\times 10^{40} \ {\rm erg}) \ B_{12}^2 R_6^3$. One can see that in order to produce an FRB-like event, one needs to dissipate a strong enough magnetic field in a large enough volume. For a favorable magnetic configuration, the requirement for pulsar parameters may be less demanding. 

Finally, EMRIs are another type of target to search for the cCBC signal. The systems with a relatively small mass ratio $q$ (within the EMRI category) and the ratio $a/a_{\rm min}$ close to unity are favorable targets to search for such a signal in the era of space gravitational wave astronomy. 

\acknowledgments
This work is partially support by a NASA Astrophysical Theory Program grant NNX15AK85G. I thank Kunihito Ioka, Kazuya Takahashi, and Tomoki Wada for discussing NS charge physics and pointing out an error in the previous version.




\end{document}